\documentclass[apl,twocolumn]{revtex4-1}

\usepackage{graphicx}
\usepackage{dcolumn}
\usepackage{bm}

\begin{document}

\preprint{AIP/123-QED}

\title{Photocurrents  in Bi$_{2}$Se$_{3}$: bulk versus surface, and injection versus shift currents}

\author{Derek A. Bas}
\affiliation{Department of Physics and Astronomy, West Virginia University, Morgantown, West Virginia 26506-6315, U.S.A.}

\author{Rodrigo A. Muniz}
\affiliation{Department of Physics and Institute of Optical Sciences, University of Toronto, Toronto, Ontario M5S 1A7, Canada}

\author{Sercan Babakiray}
\affiliation{Department of Physics and Astronomy, West Virginia University, Morgantown, West Virginia 26506-6315, U.S.A.}
\author{David Lederman}
\affiliation{Department of Physics and Astronomy, West Virginia University, Morgantown, West Virginia 26506-6315, U.S.A.}
\author{J. E. Sipe}
\affiliation{Department of Physics and Institute of Optical Sciences, University of Toronto, Toronto, Ontario M5S 1A7, Canada}
\email[]{sipe@physics.utoronto.ca}
\author{Alan D. Bristow}
\affiliation{Department of Physics and Astronomy, West Virginia University, Morgantown, West Virginia 26506-6315, U.S.A.}
\email[]{alan.bristow@mail.wvu.edu}

\date{\today}

\begin{abstract}

Optical injection and detection of charge currents can complement conventional transport and photoemission measurements without the necessity of invasive contact that may disturb the system being examined. This is a particular concern for the surface states of a topological insulator. In this work one- and two-color sources of photocurrents are examined in epitaxial, thin films of Bi$_{2}$Se$_{3}$. We demonstrate that optical excitation and terahertz detection simultaneously captures one- and two-color photocurrent contributions, as previously not required in other material systems. A method is devised to isolate the two components, and in doing so each can be related to surface or bulk excitations through symmetry. This strategy allows surface states to be examined in a model system, where they have independently been verified with angle-resolved photoemission spectroscopy.

\end{abstract}

\maketitle

\section{Introduction}

Photocurrents are a versatile tool for studying the intricacies of the charge carrier dynamics in a host of materials. Even for the simplest semiconductors, subject to excitation in the limit of long pump pulses at a single carrier frequency, there is a complex assortment of photocurrents. These "one-color effects" arise in noncentrosymmetric crystals such as GaAs. If the photon energy $\hbar \omega$ is below the band gap $E_g$, the pump pulse can induce optical rectification (OR); here in a simple description the induced dipole moment per unit volume follows the intensity of the pulse. If $\hbar \omega > E_g$, then shift currents can arise, where in a simple description the current density follows the intensity of the pulse, as the center of charge within a unit cell moves as light is absorbed and electrons are promoted from valence to conduction bands \cite{nastos_optical_2006,cote_rectification_2002}. Finally, for semiconductor crystals with a lower symmetry than GaAs, such as those that form in the wurtzite structure, excitation above the band gap can also lead to an injection current, where electrons and holes are injected preferentially on one side of the Brillouin zone; here in a simple description, neglecting scattering, the time rate of change of the injected current follows the intensity. These three processes of optical rectification, shift currents, and injection currents are all aspects of the same optical response, and the connections between them have been studied in detail for the noncentrosymmetric crystals in which they arise \cite{nastos_optical_2010}. Typically the currents due to injection are the largest if they survive, and if not the shift currents are larger than the rectification currents, but there are exceptions: Bieler et al. \cite{bieler_simultaneous_2006} found that shift currents are larger than injection currents in (110)-grown GaAs/AlGaAs quantum wells, which have reduced symmetry due to quantum confinement.

In centrosymmetric crystals, such as silicon, one-color effects do not survive. However, photocurrents can arise from the interference of light with carrier frequencies at $\omega$ and $2\omega$ when $2\hbar \omega$ crosses the gap \cite{costa_all-optical_2007}. Such "two-color" injection currents survive in any material, as do two-color shift and optical rectification currents. In fact, two-color injection currents were first studied in GaAs crystals \cite{hache_observation_1997}. The injection current is expected to dominate when $2\hbar \omega $ crosses the band gap.

Graphene has a band structure that distinguishes it from the usual semiconductors, in that its states near the six symmetric K points form Dirac cones with massless dispersion. Two-color photocurrents have been observed \cite{sun_coherent_2010}, and the expressions for the nonlinear optical response that describes them have been worked out in detail \cite{cheng_third-order_2015} for frequencies where states near the K points are important, with scattering included phenomenologically.Unlike most semiconductors, where two-color injection currents have been studied for $\hbar \omega <E_{g} < 2\hbar \omega$, in undoped graphene one-photon absorption is possible at both $2\omega $ and $\omega $, and so complexities in the injected currents arise \cite{rioux_interference_2014}. Yet since the lattice structure has inversion symmetry there are no one-color effects.

The topological insulator (TI) bismuth selenide (Bi$_{2}$Se$_{3}$) is a unique example because its band structure is a much richer system, containing both a direct band gap in the bulk and graphene-like massless conducting states near the surface. Topological surface states are conducting, spin-locked, and protected by symmetry from backscattering by nonmagnetic impurities \cite{hasan_textitcolloquium_2010,hosur_circular_2011,stevens_quantum_2003,galanakis_electrostatic_2012}. Bi$_{2}$Se$_{3}$ is the prototypical TI (albeit doped in the as-grown samples), and has been intensively studied for its electronic, thermoelectric \cite{mishra_electronic_1997,hor_$p$-type_2009}, and optical properties \cite{hsieh_nonlinear_2011,kumar_spatially_2011,di_pietro_optical_2012,valdes_aguilar_terahertz_2012,glinka_ultrafast_2013,lu_third_2013,mciver_control_2012,sim_ultrafast_2014,mciver_theoretical_2012}. It has been proposed \cite{muniz_coherent_2014} that two-color injection currents can be injected by a laser pump and its second harmonic, and this was observed \cite{bas_coherent_2015} in Bi$_{2}$Se$_{3}$ thin films using a terahertz (THz) probe.

Many surface effects come into play at interfaces between TIs and other materials, some of which can lead to exotic phenomena \cite{fu_superconducting_2008,wang_calculation_2013}, but very often they only introduce complications in experiments designed to study the topological surface states \cite{seo_transmission_2010}. This suggests that photocurrents would constitute an ideal probe, since contacts at the surface can be avoided. Due to the relatively small band gap of Bi$_{2}$Se$_{3}$ ($\sim$0.3 eV), optoelectronic applications involving explicitly the properties of the topological surface states had been thought to require optical fields with mid-to-far infrared frequencies.
However, recently a second Dirac cone of surface states (SS2) was discovered at 1.7 eV above the commonly studied Dirac point (SS1) \cite{sobota_direct_2013}. This shows that surface-to-surface optical transitions can be excited using NIR commercial solid-state lasers ($\sim$0.8 eV) that offer the possibility of extreme control of the dynamics.

Just as important for the study of a TI such as Bi$_{2}$Se$_{3}$ is that in the bulk the material is centrosymmetric, and thus no one-color effects can arise \cite{mciver_control_2012,hsieh_nonlinear_2011,olbrich_room-temperature_2014}. Nonetheless, if it is excited with light at $\omega $ and $2\omega $, with $2\hbar \omega >E_{g}$, two-color injection currents will appear. At the surface the symmetry is broken, and in fact a one-color shift current is allowed \cite{olbrich_room-temperature_2014,zhu_effect_2015,braun_ultrafast_2015}. The presence of both two-color and one-color photocurrents is both a challenge and an opportunity. The challenge is to be able to separate the two effects and identify just to what an experiment is sensitive. The opportunity is that by examining the interplay between these two photocurrents a noninvasive study of the electronic states both in the bulk and at the surface should be possible. In this article, we demonstrate a scheme for detection of these two types of photocurrents. For each photocurrent the dependence on the angle between the light polarization and the crystal lattice axis is measured and found to agree with theoretical predictions. We present a comparison of the strengths of the shift and injection currents, and outline a method for isolating the injection current using simple measurements of the total two-color photocurrent.

\section{Experiment}

\begin{figure}[htb]
\centering
\includegraphics[width=\linewidth]{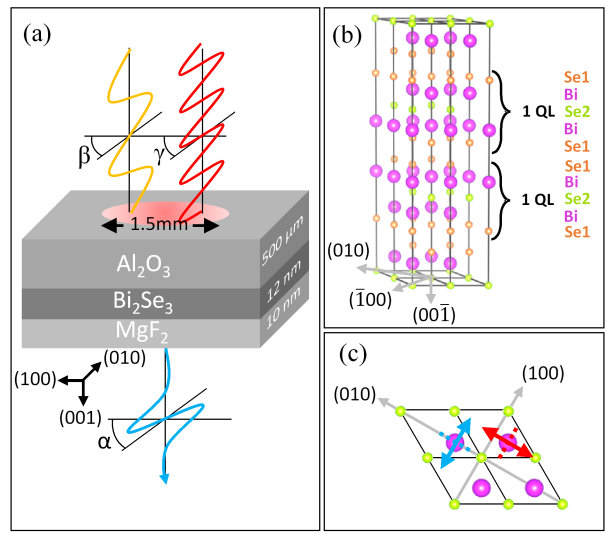}
\caption{Schematic diagram of experimental configuration. (a) Here $\beta$ and $\gamma$ are the angles that the $\omega$ and $2\omega$ pump polarizations make with the (100) axis of the TI; in the experiments reported here $\beta=\gamma$. The angle of the polarization of the detected THz relative to the (100) axis of the TI is denoted by $\alpha$; in this work $\alpha=\beta$ corresponds to the polarization of the detected THz parallel to the polarization of the incident fields, and $\alpha=\beta+\pi/2$ corresponds to the polarization of the detected THz perpendicular to the polarization of the incident fields. (b) Four unit cells, showing the trigonal symmetry of the Bi$_2$Se$_3$ lattice. (c) Top-down view of the top two atomic layers shown in (b), with example pump polarizations. The blue is a symmetry plane for which the lattice looks identical in both directions. For red, the symmetry is broken and a directional shift current can occur. For $\alpha=\beta+\pi/2$, the relevant symmetry plane is along the direction of the polarization vector.}
\label{fig:Fig1}
\end{figure}

Details of sample growth can be found in D. A. Bas et al. \cite{bas_coherent_2015} and P. Tabor et al. \cite{tabor_molecular_2011}. In Fig.~\ref{fig:Fig1}(a) we show an experimental sample consisting of three distinct materials: A 500-$\mu$m sapphire substrate which is transparent to the pump radiation; a 12-nm TI grown via molecular beam epitaxy; and a 10-nm MgF$_2$ capping layer added to prevent exposure to atmosphere and inhibit the build-up of space charge fields caused by band bending near the surface \cite{zhu_effect_2015,braun_ultrafast_2015}. In the TI, bismuth (Bi) and selenium (Se) naturally grow in layers as shown in Fig.~\ref{fig:Fig1}(b), which shows four rhombohedral unit cells. Bi layers are always surrounded by Se layers, but Se layers take two unique positions: Se2 layers are surrounded by Bi, but Se1 layers are separated by van der Waals bonds on one side with other Se1 layers. This van der Waals interface occurs every five atomic layers, and causes samples to naturally grow in whole number quintuple layer (QL) intervals, with an Se1 layer on the end. In Bi$_2$Se$_3$, 1~QL is approximately 1~nm thick.
The layered structure makes the material applicable for exfoliation, which has been used in many other studies.

The experimental setup \cite{bas_coherent_2015} uses a laser amplifier system with an optical parametric amplifier (OPA) to provide $\sim$80-fs pulses at a repletion rate of 1~kHz. Signal pulse from the OPA centered at 1540 nm (0.8 eV) pump a $\beta$-barium borate crystal to generate second-harmonic pulses at 770~nm (1.6~eV). These fundamental ($\omega$) and frequency-doubled ($2\omega$) pulses feed a two-color Mach-Zehnder interferometer that independently controls the intensity, phase, and polarization of the two pulses. The combined pulses impinge at normal incidence the Bi$_2$Se$_3$ sample which sits in a rotation mount. The acceleration of charge due to the production and decay of the photocurrent leads to THz radiation, which is collected using off-axis parabolic mirrors and detected with electro-optic sampling (EOS). The 80-fs gate pulses used for EOS are derived from the laser amplifier. By varying the delay time $t_{det}$ of the gate pulse the THz signal is mapped out in the time domain. The time of maximum field detected resulting from the $\omega$ pump alone is indicated by $t_{det}=0$. The detection is set up to measure the components of the THz radiation polarized parallel and perpendicular to the polarization of the incident radiation.

The powers of the pulse trains at $\omega$ and $2\omega$ were approximately 35~mW and 1~mW, respectively. Both beam diameters were collimated to have a $1/e^2$ diameter of approximately 1.5~mm, and were co-linearly polarized in a fixed direction with the azimuthal rotation of the sample controlled by a rotation mount. Fig.~\ref{fig:Fig1}(a) illustrates a generalized case where $\beta$ and $\gamma$ are the angles that the $\omega$ and $2\omega$ pumps make with the (100) axis of the TI, but throughout this work $\beta=\gamma$. When the (100) axis of the TI is parallel with the pump polarization, $\beta$ is set to zero. Lattice orientation for the sample was confirmed by x-ray diffraction measurements; however, a distinction is not made between (100) and ($\bar{1}$00) directions. We use $\alpha$ to denote the polarization of the measured THz field, and for this work orthogonal components at $\alpha=\beta$ (parallel) and $\alpha=\beta+\pi/2$ (perpendicular) were recorded. Detection was calibrated using a strong OR source \cite{rowley_broadband_2012}. The $\omega$ and $2\omega$ pulses transmitted first through the substrate and then the TI to avoid stretching of the THz emission by the substrate \cite{grischkowsky_far-infrared_1990}.
Optical phase walk-off of the pump pulses was corrected by the two-color interferometer.

The path length of the $\omega$ pulse was fixed throughout the entire experiment. The delay time $\tau$ of the $2\omega$ pulse (and hence the phase parameter $\Delta\phi=\phi_{2\omega}-2\phi_{\omega}$) was controlled by a pair of glass wedges and $\tau=0$ was identified with the point of maximum injection current within the cross-correlation of the two excitation pulses. By adjusting $\tau$ beyond the cross-correlation envelope, one might expect the two-color injection current to be completely suppressed, allowing the possibility of independently measuring a shift current and comparing it to the injection current. However, it has been shown before \cite{ruppert_ultrafast_2012}, and confirmed below, that this is not the case; an individual shift current can be reliably measured only by blocking the second pump.

\section{Results and discussion}

\subsection{Shift and injection currents}

A shift current can be induced by a single optical pump field $\hbar\omega>E_g$ or $2\hbar\omega>E_g$.
The charge distribution in real space associated with a conduction band state is usually displaced from the charge distribution associated with the valence band state at the same position in the Brillouin zone; one charge distribution might be centered on the Bi atoms, and the other on the Se atoms, for example. During absorption, it is then possible for the charge distribution to evolve from that of the valence band state to that of the conduction band state in such a way that there is a net current associated with the change of state. This occurs most typically if the pump field vector points from one atom to the neighboring atom (red vector in Fig.~\ref{fig:Fig1}(c)). In the other polarization shown by the blue vector, there is no preference for the direction of the charge motion and no net shift current results.

The lowest order contribution to the shift current is governed by a third rank tensor, and associated with a divergent part of $\chi^{(2)}$ \cite{nastos_optical_2006,nastos_optical_2010}. It can be computed in terms of the incident field $E_\omega$ through a third-rank tensor $\nu_{abc}$ given by 
\begin{equation}
J_{a}^{shift} = \nu_{abc} E_{-\omega}^{b}E_{\omega}^{c} +c.c.
\label{eq:power}
\end{equation}
The tensor $\nu_{abc}$ satisfies the same symmetries as the crystal lattice. It vanishes for centrosymmetric materials such as Bi$_2$Se$_3$, but inversion symmetry is broken at the surface and $\nu_{abc}$ can have non-vanishing components. For surface states in Bi$_2$Se$_3$, the only non-zero tensor component is $\nu_{xxx}$. The parallel $\left(\alpha=\beta\right)$ component of the current $J_{\parallel}$, and the perpendicular $\left(\alpha=\beta+\frac{\pi}{2}\right)$ component $J_{\perp}$ can be written as
\begin{equation}
\begin{array}{rl}
J^{shift}_{\parallel}= & \nu_{xxx}\cos\left(3\beta\right)\left|E_{\omega}\right|^{2}, \\
J^{shift}_{\perp} = & -\nu_{xxx}\sin\left(3\beta\right)\left|E_{\omega}\right|^{2}.
\end{array}
\label{eq:shift}
\end{equation}
Similar terms exist for excitation with $|E_{2\omega}|^{2}$.   

To induce an injection current one must use two phase-related pumps which overlap in time and space, which can have energies $\hbar\omega$ and $2\hbar\omega$, and can be described by the phase parameter $\Delta\phi$. Injection currents occur as a result of quantum interference between one- and two-photon absorption pathways creating an interference pattern in momentum space, and hence an asymmetric distribution of excited electrons. The lowest order contribution to the injection current is associated with a divergent part of $\chi^{(3)}$, and governed by a fourth rank tensor $\eta_{abcd}$ given by
\begin{equation}
\frac{d}{dt}J_{a}^{inj}= \eta_{abcd}E_{-\omega}^{b}E_{-\omega}^{c}E_{2\omega}^{d}e^{i\Delta\phi}+c.c.
\end{equation} 
For both the surface and bulk of Bi$_2$Se$_3$, there are three independent components of $\eta$, namely $\eta_{xxxx}$, $\eta_{xyyx}$ and $\eta_{xyxy}$. The injection rate of the current along the direction of the incident field $\left(\alpha=\beta\right)$ is 
\begin{equation}
\begin{array}{rl}
\frac{d}{dt}J_{\parallel}^{inj}= &  2[\textrm{Re}\left(\eta_{xxxx}\right)\cos\left(\Delta\phi\right) \\
& +  \textrm{Im}\left(\eta_{xxxx}\right)\sin\left(\Delta\phi\right)]\left|E_{\omega}\right|^{2}\left|E_{2\omega}\right|,
\end{array}
\label{eq:injdir}
\end{equation}
while the component of the injection current perpendicular to the incident field $\left(\alpha=\beta+\frac{\pi}{2}\right)$ vanishes $\frac{d}{dt}J_{\perp}^{inj}=0$.

\subsection{Observation of competing current sources}

\begin{figure}[h!]
\centering
\includegraphics[width=\linewidth]{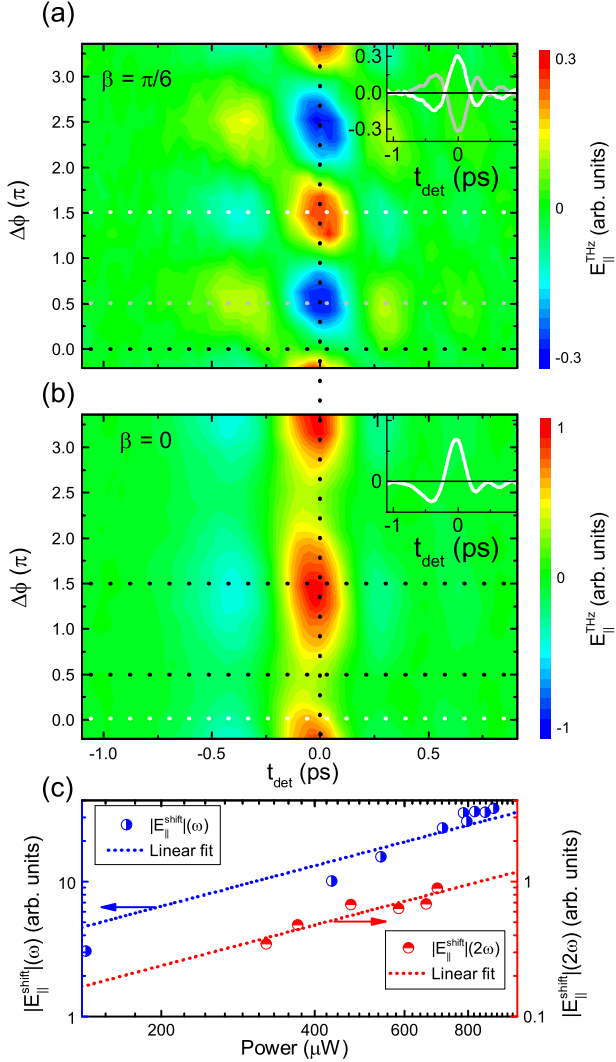}
\caption{THz electric field emitted by Bi$_{2}$Se$_{3}$ as a function of $t_{det}$ and $\Delta\phi$.
The dotted vertical line indicates $t_{det}=0$, the maximum for which the later polar angle measurements are taken. Dotted horizontal lines indicate transients for $\Delta\phi=0$, $\pi/2$, and $3\pi/2$. (a) Here $\beta=\pi/6$, resulting in a vanishing shift current so only the characteristic checkerboard pattern of the injection current is visible. The gray and white lines in the inset indicate the evolution of the THz electric fields that follow from the injection currents at $\Delta\phi=\pi/2$ and $3\pi/2$. (b) Here $\beta=0$, resulting in a maximum positive shift current offsetting the oscillating injection current. The white line shown in the inset indicates the evolution of the THz electric field following from the shift current. (c) Power dependence of the electric field produced by a single pulse $\omega$ (left) or $2\omega$ (right).}
\label{fig:Fig2}
\end{figure}

Fig.~\ref{fig:Fig2} shows the THz emission versus $t_{det}$ and $\Delta\phi$ in parallel detection configuration ($\alpha=\beta$). In (a), $\beta=\pi/6$ results in suppression of the shift current. With the sample rotation at $\beta=\pi/6+n\pi/3$ (where $n$ is an integer) no signal at $\alpha=\beta$ occurs for a one-color excitation. The observed grid pattern in this case is similar to other injection current experiments on GaAs, silicon, and graphene \cite{hache_observation_1997,costa_all-optical_2007,sames_all-optical_2009,sun_coherent_2010}. The THz transients shown in the inset are averages of the horizontal lines at $\Delta \phi = 3\pi/2 + 2\pi n$ (white) and at $\Delta \phi = \pi/2 + 2\pi n$ (gray). They follow from the dynamics of the injection current alone. 

In Fig.~\ref{fig:Fig2}(b) $\beta=0$ and shift and injection currents are simultaneously observed. For $\beta = 0+n\pi/3$ radians, maximum shift currents occur along $\alpha=\beta$ because a neighboring atom is aligned with the pump polarization. This behavior follows from Eq.~\ref{eq:shift}. The THz transient shown in the inset is an average of all of the horizontal lines from $\Delta \phi = -\pi/2$ to $\Delta \phi = 7\pi/2$. Averaging over a multiple of $2\pi$ removes the oscillating injection current, and therefore the white line follows from the dynamics of the shift current alone. It contains contributions from both the pump at $\omega$ and that at $2\omega$.

In the far field, the idealized shape of the THz electric field follows the form of $dJ/dt$ \cite{cote_simple_2003}. In model systems, this leads to different shaped transients for the injection and shift current. In contrast, even epitaxial growth of Bi$_{2}$Se$_{3}$ does not result in the material quality of lattice-matched semiconductors such as GaAs or Si. Scattering can reduce the injection current strength and longevity leading to similar shaped transients. Indeed, the magnitudes of injection and the $2\omega$ shift currents are very similar too, which indicates that the surface has a strong response, as has been observed in graphene, or that the bulk is somehow suppressed.

The power dependence of the shift currents is shown in Fig.~\ref{fig:Fig2}(c). The THz electric field (and hence the current) for both $\omega$ and $2\omega$ excitation is linear in the pump power, as expected from Eq.~\ref{eq:power}. The $2\omega$ shift current is approximately an order of magnitude weaker than that for $\omega$ excitation, due mainly to the excitation irradiance at the two pulse energies, and the difference in optical absorption coefficient, which are $3\times 10^{4}$ cm$^{-1}$ for $\omega$ and $1\times 10^{6}$ cm$^{-1}$ for $2\omega$.

In addition to shift and injection currents, other effects such as photon drag, where momentum is transferred between photons and electrons via scattering, can play a role in the generation of photocurrents. Previous studies have taken care to demonstrate that such effects are negligible in Bi$_{2}$Se$_{3}$ \cite{mciver_control_2012,olbrich_room-temperature_2014}.

\begin{figure}[h!]
\centering
\includegraphics[width=\linewidth]{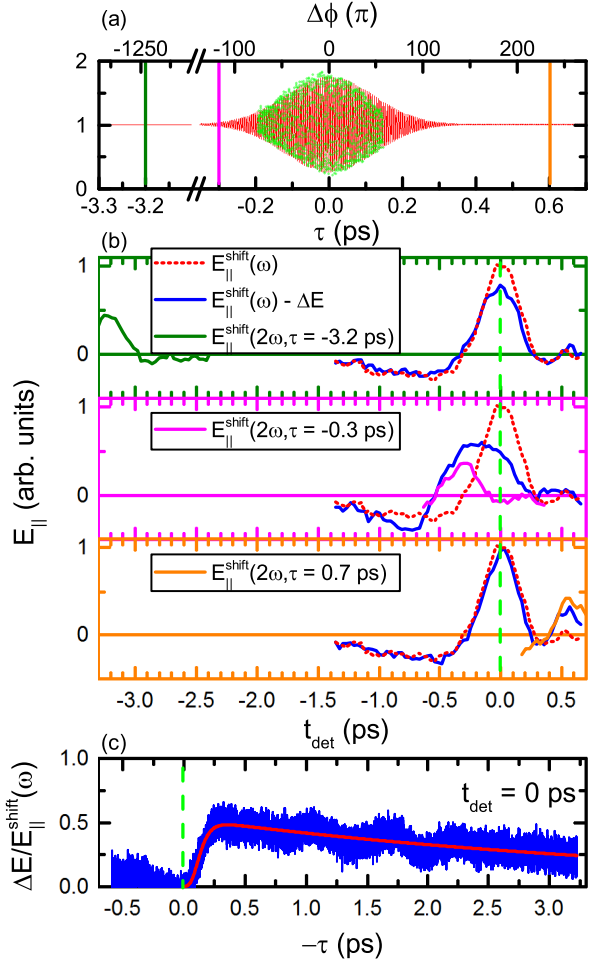}
\caption{Measured electric field from shift currents as a function of $\tau$. (a) The range of $\tau$ values with best overlap and maximum injection current occurring at $\tau=0$, with vertical lines indicating the $2\omega$ pulse occurring before, at the onset, and after the $\omega$ pulse, conditions shown in (b). The dotted red lines in (b) show $E_{||}^{shift}(\omega)$, the THz electric field resulting from the shift current, measured when only $\omega$ is present. The blue lines in (b) show how the field is modified by the presence of $2\omega$ at three delay times. This behavior at fixed time $t_{det}=0$ over the entire range of $\tau$ values is shown in (c).}
\label{fig:Fig3}
\end{figure}

It is observed that even when $\tau$ is much too great for two-color injection current to occur, the signal observed at $t_{det}=0$ is modified by the presence of the $2\omega$ pulse. Fig.~\ref{fig:Fig3} shows this phenomenon. Fig.~\ref{fig:Fig3}(a) shows the phase-dependent cross-correlation of the injection current offset by the shift current as green data, with a fit of the form $A\exp\left[-\frac{\tau^2}{0.36(\tau_{x})^2}\right]\sin\left[\omega(\tau-\tau_{0})\right]+1$ shown in red, where the FWHM of the cross-correlation is $\tau_{x}=0.28$~ps, $\tau_{0}$ is an arbitrary phase shift, and $A$ represents the relative strength of the injection current. Also shown are three positions of the $2\omega$ pulse outside the current injection envelope, used below. When the $2\omega$ pump is blocked, the $E^{shift}_{||}(\omega)$ contribution can be seen as the dotted red line in Fig.~\ref{fig:Fig3}(b). However, this contribution is modified by the $2\omega$ pump, even when it impinges the sample a full 3.2~ps ($\sim$40 pulse widths) before the $\omega$ pump. The THz signal observed is modified by approximately 20\% in this case, and the modification increases steadily to 50\% as $\tau$ becomes closer to 0 ps. This is probably due to free carrier absorption of the THz \cite{harrel_influence_2010} and indicates that a trivial solution of simply adding the two separate shift currents together is not necessarily appropriate. In Fig.~\ref{fig:Fig3}(c), this effective optical pump and THz probe contribution is mapped out over the entire available range of $\tau$. The transient signal is fit with a pulse model of the form $A\left[1-\exp\left(-\frac{\tau-\tau_{0}}{\tau_1}\right)\right]^P \exp\left[-\frac{\tau-\tau_{0}}{\tau_2}\right]$, with rise time $\tau_1=0.06$ ps and decay time $\tau_2=4.09$ ps; $A$ and $P$ are arbitrary fitting parameters to match the signal strength and $\tau_{0}$ is an arbitrary temporal shift.
Any signal occurring at $\tau>0$ was ignored for this fit.

\subsection{Separation of shift and injection currents}
\label{sec:isolation analysis}

\begin{figure}[h!]
\centering
\includegraphics[width=\linewidth]{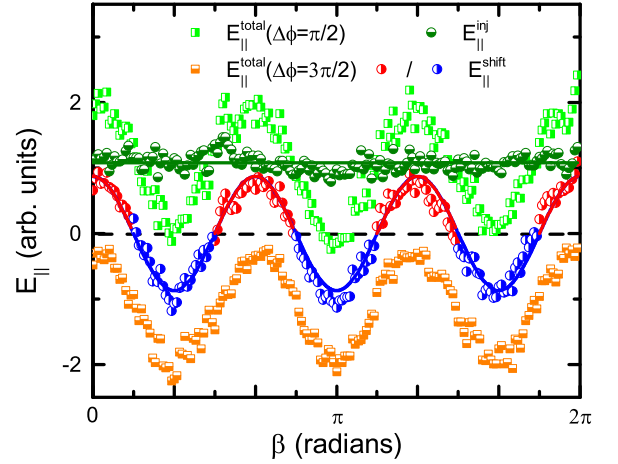}
\caption{Demonstration of current separation analysis based on Eq.~\ref{eq:inj}. Light green is the two-color photocurrent at $\Delta\phi=\pi/2$ and orange is the same at $\Delta\phi=3\pi/2$. Subtraction gives the injection current in olive, and addition gives the shift current in red/blue. Solid lines for $E^{shift}$ and $E^{inj}$ are theoretical models fit to experimental amplitude.}
\label{fig:Fig4}
\end{figure}

The total current measured by the THz probe is a sum of both one-color shift currents and the two-color injection currents. Injection currents depend on the phase parameter $\Delta \phi$ and shift currents are independent of it. Since $J_{\parallel}^{inj} \left(\Delta\phi=\frac{\pi}{2}\right) = -J_{\parallel}^{inj} \left(\Delta\phi=\frac{3\pi}{2}\right)$, if the total current is measured for both values of the phase parameter 
\begin{equation}
\begin{array}{rl}
J_{\parallel}^{total} \left(\Delta\phi=\frac{\pi}{2}\right) = & J_{\parallel}^{shift} +J_{\parallel}^{inj} \left(\Delta\phi=\frac{\pi}{2}\right), \\
J_{\parallel}^{total} \left(\Delta\phi=\frac{3\pi}{2}\right) = & 
J_{\parallel}^{shift}+J_{\parallel}^{inj} \left(\Delta\phi=\frac{3\pi}{2}\right),
\end{array}
\end{equation}
it is possible to extract both the injection and shift currents as  
\begin{equation}
\begin{array}{rl}
J_{\parallel}^{inj} \left(\frac{\pi}{2}\right) 
= &
\frac{1}{2} \left[ J_{\parallel}^{total}\left(\frac{\pi}{2}\right) -J_{\parallel}^{total} \left(\frac{3\pi}{2}\right) \right], \\
J_{\parallel}^{shift} = &
\frac{1}{2} \left[ J_{\parallel}^{total}\left(\frac{\pi}{2}\right) +J_{\parallel}^{total} \left(\frac{3\pi}{2}\right) \right].
\end{array}
\label{eq:inj}
\end{equation}
In Fig.~\ref{fig:Fig4} this treatment is demonstrated, showing that the injection current is indeed $\beta$-independent, while the sum of the shift currents follows the same trend as the individual shift currents.

\begin{figure}[h!]
\centering
\includegraphics{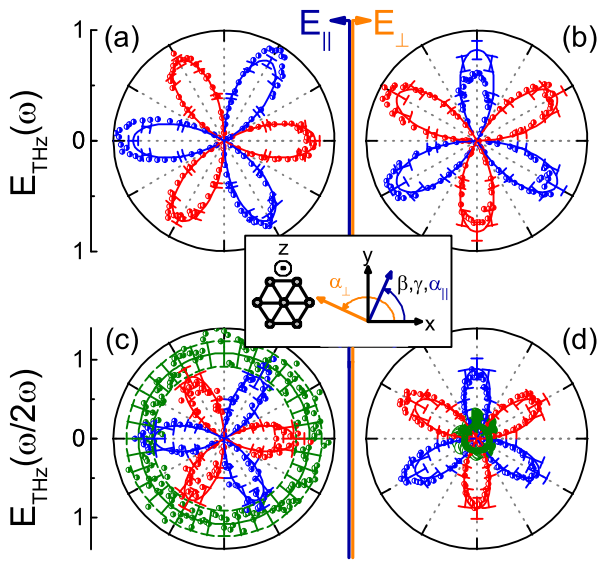}
\caption{THz traces as a function of $\beta$, all obtained at time $t_{det}=0$. Inset shows the relevant angles.
Red and blue lines in (a) - (d) indicate positive and negative values, respectively.
In (a) and (c), the parallel ($\alpha=\beta$) radiation was detected, and in (b) and (d), the perpendicular ($\alpha=\beta+\pi/2$) was detected.
In (a) and (b), we see the current produced by the $\omega$ pump alone.
In (c) and (d), both pumps were incident simultaneously and the shift and injection current contributions were calculated using Eq.~\ref{eq:inj}.
Error bars (15\% maximum for two-color measurements) were obtained by repeating scans multiple times and recording the difference in amplitudes for the sinusoidal fits.
All data have been normalized to the magnitude of $E^{shift}_{||}(\omega)$.
Solid lines are theoretical models with magnitudes fitted to the data where shift currents are assumed to be equal in parallel and perpendicular detection, and interference current is assumed to be zero in perpendicular detection.}
\label{fig:Fig5}
\end{figure}

The same $\beta$-dependence stemming from the lattice symmetry is reproduced by all shift currents measured (see Fig.~\ref{fig:Fig5}). The peak amplitudes of the shift currents with polarization parallel to the polarization of the $\omega$ and $2\omega$ pulses can each be fit well by a cosine of the form $A\cos[2\pi\beta/T]$ where $T=2\pi/3$, shown as solid lines in the figure. The peak amplitudes of the shift currents with polarization perpendicular to the polarization of the $\omega$ and $2\omega$ pulses are fit using $A\cos[2\pi(\beta-\frac{\pi}{6})/T]$, following the relationship in Eq.~\ref{eq:shift}. The difference in magnitude between the parallel and perpendicular contributions is negligible within experimental uncertainty.

In Fig.~\ref{fig:Fig5}(c) and (d) the shift current and injection current contributions are calculated using $\omega+2\omega$ excitations (Eq.~\ref{eq:inj}). Azimuthal angle measurements were recorded for conditions of minimum and maximum injection current ($\Delta\phi=\pi/2$ and $3\pi/2$, respectively). To minimize the effects of phase drift (which was observed to be as low as one cycle per hour), measurements were recorded for $2\pi/3$ radians at a time, and the phase was re-optimized in between before continuing. After subtraction to obtain the injection current contribution, the results were found to be approximately independent of angle, with a residual 4.2\% anisotropy (fit amplitude/offset).

Going from the detection of THz radiation with polarization parallel to that of the incident pulses to that with polarization perpendicular to that of the incident pulses, we see that the shift current pattern is rotated by an angle of $\pi/6$, and the magnitude is increased by only $\left(\frac{E_{\parallel}-E_{\perp}}{E_{\parallel}+E_{\perp}}\right)=1.8\%$, well within experimental uncertainty. We also see a reduction in the injection current of approximately 74\%. As described by Eq.~\ref{eq:injdir}, injection currents are typically along the direction of the pump polarization. The reduction trends toward the expected value of 100\%. Currents perpendicular to the pump polarization may arise due to the strong spin-orbit coupling in the material, as has been observed in other semiconductors \cite{zhao_coherence_2006}.

It is known that z-direction components will arise from non-normal excitation, which yields further interesting possibilities for the generation of photocurrents \cite{zhu_effect_2015,braun_ultrafast_2015}. Their generation may also result due to the built-in electric field caused by band-bending near the surfaces \cite{bahramy_emergent_2012}. The experiments performed here were unable to detect currents traveling in the z-direction, but their presence cannot be entirely ruled out. Nevertheless, the strong shift current at $2\omega$ relies on surface (such as SS1-to-SS2 \cite{sobota_direct_2013}) transitions and the strength of this shift current is comparable to the two-color injection current that is allowed in the bulk. Surface SS1-to-SS2 contributions, therefore, have comparable photocurrent strength to those from bulk which has a larger density of states in the bulk \cite{bas_coherent_2015} and an expected number of joint density of states. Since the bulk photocurrents may be subject to more scattering, this meaning that the topological states most likely have a higher mobility and are subject to lower back scattering. At $\omega$, the shift current is weaker and this photon energy does not correspond to SS1-to-SS2 transitions, although must still include a surface state. Hence, isolation of the shift and injection current indeed provide information about the surface and bulk contributions.

\section{Conclusions}
\label{sec:conclusion}

Photocurrents are often used as tools for exploring interesting optoelectronic properties of semiconductors. In many systems, selection rules and sample geometry preclude some photocurrent mechanisms while allowing others. Although in certain cases both one-color and two-color photocurrents have been observed simultaneously, to date they have not been isolated and studied in-depth in a rich system like the distinctive bulk and surface state energy bands in three-dimensional topological insulators. Adding and subtracting currents taken from separate one- and two-color measurements was shown to have limitations coming from interplay between the two pulses even outside the cross-correlation envelope. Here, an all-optical coherent control method is exploited to isolate the total shift current arising from two one-color excitations and the two-color injection current generated through quantum interference. It is shown that a relative-phase and polarization dependence analysis of the two-color measurement allows for extraction of the total shift current.

Comparison of the total shift and injection currents with the individual shift currents, detected via terahertz emission, and compared to theoretical predictions gives information about the origins of each photocurrent. The dependence on the direction of the polarization of the incident fields confirms that the shift currents arises from surface-to-surface transitions where possible and is comparable to the bulk injection that would be expected to be stronger in a near-perfect material. This strategy has the potential to provide great insight into the optoelectronic properties of Bi$_2$Se$_3$, where topological surface states are known to exist. Therefore, this powerful tool for isolating the photocurrents can provide a promising bridge between angle-resolved photoemission spectroscopy and transport measurements in similarly rich materials, where topological surface states have not yet been confirmed, or in materials that also exhibit generation of multiple and simultaneous photocurrents and in devices where non-invasive measurement is required.

\section*{Acknowledgments}
The authors wish to thank Tudor Stanescu for useful discussions, and Trent Johnson and Pavel Borisov for help with sample preparation.

\section*{Funding Information}
WV Higher Education Policy Commission (HEPC.dsr.12.29); 
National Science Foundation (CBET-1233795);
Natural Sciences and Engineering Research Council of Canada (NSERC).

\end{document}